# Emergence of Chern metal in a moiré Kondo lattice

Wenjin Zhao[1*†], Zui Tao[2*], Yichi Zhang[3*], Bowen Shen[2], Zhongdong Han[3], Patrick Knüppel[3], Yihang Zeng[3], Zhengchao Xia[2], Kenji Watanabe[4], Takashi Taniguchi[4], Jie Shan[1,2,3,5†], Kin Fai Mak[1,2,3,5†]

[1]Kavli Institute at Cornell for Nanoscale Science, Ithaca, NY, USA
[2]School of Applied and Engineering Physics, Cornell University, Ithaca, NY, USA
[3]Laboratory of Atomic and Solid State Physics, Cornell University, Ithaca, NY, USA
[4]National Institute for Materials Science, 1-1 Namiki, 305-0044 Tsukuba, Japan
[5]Max Planck Institute for the Structure and Dynamics of Matter, Hamburg, Germany.

*These authors contributed equally.
†Email: wz435@cornell.edu; jie.shan@cornell.edu; kinfai.mak@cornell.edu

**Abstract**
A Chern metal is a two-dimensional metallic state of matter carrying chiral edge states. It can emerge as a doped Chern insulator, but theoretical studies have also predicted its emergence near a Kondo breakdown separating a metallic chiral spin liquid and a heavy Fermi liquid in a frustrated lattice. To date, the latter exotic scenario has not been realized. Here, we report the observation of a Chern metal at the onset of the magnetic Kondo breakdown in a frustrated moiré Kondo lattice--angle-aligned $MoTe_2$/$WSe_2$ bilayers. The state is compressible and is manifested by a nearly quantized Hall resistance but a finite longitudinal resistance that arises from a bad metallic bulk. The state also separates an itinerant and a heavy Fermi liquid and appears far away from the band inversion critical point of the material, thus ruling out its origin from simply doping a Chern insulator. We demonstrate the presence of a chiral edge state by nonlocal transport measurements and current-induced quantum anomalous Hall breakdown. Magnetic circular dichroism measurements further reveal a magnetization plateau for the Chern metal before a metamagnetic transition at the Kondo breakdown. Our results open an opportunity for moiré engineering of exotic quantum phases of matter through the close interplay between band topology and Kondo interactions.

**Main**
Moiré materials have emerged as a highly tunable platform for the realization of strongly correlated phases of matter(*1-5*). In addition to emulating the Hubbard model, they can be engineered to emulate the Kondo lattice model(*6-18*), in which a lattice of local magnetic moments interacts with a sea of itinerant electrons through the Kondo interaction. Compared to bulk Kondo materials, the tunability and the widely available topological flat bands in moiré materials open an opportunity to explore Kondo lattice physics in new regimes, as exemplified by proposals ranging from topological heavy fermions in twisted bilayer(*8*) and trilayer(*7*) graphene, to a chiral Kondo lattice in semiconducting transition metal dichalcogenide (TMD) heterobilayers(*9, 11, 12, 14-17*), and to a topological Kondo lattice in twisted bilayer $WSe_2$ (Ref. (*18*)). Specifically, the highly gate-tunable Kondo interactions in TMD moiré Kondo lattices make it possible to study the breakdown of Kondo singlets in a continuous manner(*6, 9, 10, 12-14*). A transition between a heavy

Fermi liquid and a metallic chiral spin liquid with an intermediate Chern metal phase has been predicted in the presence of geometric frustrations(9, 19). The Chern metal concurrently hosts a metallic bulk and a chiral edge state and is intimately connected to the chiral spin liquid. Related Chern metal phases have also been proposed to promote chiral superconductivity in frustrated Kagome metals(20, 21). However, experimental evidence of a Chern metal beyond simply doping a Chern insulator(22) has yet remained elusive.

Here we demonstrate the emergence of a Chern metal near the magnetic Kondo breakdown in angle-aligned MoTe$_2$/WSe$_2$ moiré bilayers(10-17, 23). The 7% lattice mismatch between MoTe$_2$ and WSe$_2$ (Mo and W, respectively, for short) produces a moiré period of $a_M \approx 5$ nm and a moiré density of $n_M \approx 5 \times 10^{12}$ cm$^{-2}$ (Ref. (10, 13, 23-25)). Density functional theory calculations have established that for the first moiré valence bands originated from the K/K' states of the TMDs, the Mo- and W-Wannier orbitals occupy the two sublattice sites of a honeycomb (Fig. 1a, b), and the Mo-band is flat and the W-band is more dispersive(26, 27). An electric field $E$ perpendicular to the sample plane continuously tunes the sublattice potential, thereby the energy separation between the Mo- and W-bands (Fig. 1c)(10, 12-14, 23, 25-29). At low electric fields $E < E_1$, holes are first doped into the Mo-band(26, 27); the low-energy physics can be mapped to a triangular lattice Hubbard model(30-32); and the system is a Mott insulator at lattice filling factor $\nu = 1$ with each A site populated by exactly one hole. For $E_1 < E < E_2$, the W-valence band maximum lies between the Mo-lower and -upper Hubbard bands. Doping above $\nu = 1$ begins to populate the B sublattice site in the W-layer (26, 27); the Mott gap has been shown to survive(10, 13) and the system is a doped charge-transfer insulator(33). The low-energy physics can be mapped to a Kondo lattice model with an effective exchange interaction ($J$) between the local moments in the Mo-layer, a complex intralayer hopping ($t$) for the itinerant holes in the W-layer, and a Kondo exchange coupling ($J_K$) between the local moments and the itinerant holes(9, 11, 12, 14-17). At high electric fields $E > E_2$, the W-valence band inverts with the Mo-lower Hubbard band; a Chern insulator emerges at $\nu = 1$ near the band inversion critical point(23, 27, 28).

Figure 1d is the schematic phase diagram of the system in $E$ and $\nu$. The red dashed lines enclose the Kondo lattice region with total filling $\nu = 1 + x$, where the Mo-layer has fixed filling one and the W-layer has filling $x < 1$. The region spans a finite electric field range that is proportional to the Mott gap size in the Mo-layer(9, 10, 14). Previous studies have reported a heavy Fermi liquid in this region with a large Fermi surface below the Kondo coherence temperature $T^*$ (Ref. (10, 13)), which is dependent on $x$ and $E$ (Fig. S1). A perpendicular Zeeman field $B_\perp$ that exceeds the energy scale $k_B T^*$ ($k_B$ is the Boltzmann constant) can induce a Kondo breakdown from a heavy Fermi liquid (of density $1 + x$) to an itinerant Fermi liquid (of density $x$) decoupled from the local moments(10, 34). The phase boundary has been shown to take the shape of an arc (green line in Fig. 1d) under a constant magnetic field because the effective Kondo coupling is the weakest at the center of the Kondo lattice region(9, 14). In addition, a Chern metal phase (orange region) has been predicted in the vicinity of the Kondo breakdown arc(9); it separates the itinerant and the heavy Fermi liquids and occurs over an extended range ($E_1 < E < E_2$) far beyond the band inversion critical point at $E = E_2$.

**Emergence of a Chern metal**

We perform four-terminal transport measurements on dual-gated Hall bar devices of angle-aligned MoTe$_2$/WSe$_2$ moiré bilayers (Fig. 1a and Supplementary Materials). The two gates allow independent tuning of $v$ and $E$. Platinum electrodes are used to make ohmic contacts to the samples(23, 24). Unless otherwise specified, presented results are from device 1 at (lattice) temperature $T = 10$ mK. (A total of four devices has been examined in this study; the results are highly reproducible.) Figure 2a shows the longitudinal resistance ($R_{xx}$, left) and Hall resistance ($R_{xy}$, right) as a function of $v$ and $E$ for $B_\perp = 0$ T (the inset illustrates the measurement geometry). Figure 2b is the corresponding result for $B_\perp = 3$ T. The grey-shaded regions are either beyond the gate limits or too resistive for reliable four-terminal measurements. The Kondo lattice region is between the red dashed lines(10). At $B_\perp = 0$ T, we observe a quantum anomalous Hall effect(23, 35)—a quantized $R_{xy}$ accompanied by a vanishing $R_{xx}$—at $v = 1$ and $E_2 \approx 0.66$ V/nm right after band inversion. This is the above-mentioned Chern insulator. In the Kondo lattice region, we observe a heavy Fermi liquid(10) and a ferromagnetic Anderson insulator(13) for $x$ above and below $0.04 - 0.1$ (depending on $E$), respectively, also in agreement with earlier reports. Most interestingly, we observe at $B_\perp = 3$ T an arc-shaped region for the Kondo lattice that exhibits a large $R_{xy}$ accompanied by a $R_{xx}$ dip. With increasing $B_\perp$, the arc extends to higher fillings (Fig. S2).

We examine $R_{xx}$ (left) and $R_{xy}$ (right) as a function of $x$ and $B_\perp$ in Fig. 2c for a representative electric field within the Kondo lattice region, as denoted by the black dashed line in Fig. 2b. (See Fig. S3 for similar data at a lower electric field.) The Hall resistance changes sign near a critical field $B_{\perp\text{peak}}$ (star symbols in Fig. 2c), beyond which a Landau fan emerges immediately (see below for detailed discussions on $B_{\perp\text{peak}}$). This is consistent with a magnetic Kondo breakdown, across which both the size of the Fermi surface and the type and mass of the charge carriers change abruptly(10). Slightly below $B_{\perp\text{peak}}$, we observe a $R_{xy}$ hot spot accompanied by a $R_{xx}$ dip in the region of $B_\perp \approx 0.5 - 4$ T and $x \approx 0.05 - 0.2$. Unlike the Landau levels, the feature does not disperse linearly with the magnetic field; it disperses towards higher filling factors much more rapidly than the first several Landau levels.

Figure 2d shows a line cut of Fig. 2c at $x = 0.08$ (left) and the corresponding longitudinal and Hall conductivities ($\sigma_{xx}$ and $\sigma_{xy}$, right). As $B_\perp$ increases, the Hall response increases sharply and shows a nearly quantized plateau ($R_{xy} \approx 24$ k$\Omega$) over a field ranging from 0.5 – 3 T; concurrently, $R_{xx}$ dips down to about 2 k$\Omega$, corresponding to a bad metallic bulk with $\sigma_{xx} \approx \frac{e^2}{4h} < \frac{e^2}{h}$ (here $h$ and $e$ denote the Planck's constant and the elementary charge, respectively.)

We examine the temperature dependence of the new state under a representative magnetic field of $B_\perp = 2$ T in Fig. 3. Figures 3a and 3b illustrate that the peak in $R_{xy}$ near $x = 0.08$, accompanied by the dip in $R_{xx}$, is discernable below about 1.4 K. The detailed temperature dependence of $R_{xy}$ and $R_{xx}$ at $x = 0.08$ shows that the temperature scale of the state is about 1 K, below which $R_{xy}$ and $R_{xx}$ plateau at about 24 k$\Omega$ and 2 k$\Omega$, respectively (Fig.

3c, top). In contrast, the temperature scale of the Chern insulator ($\nu = 1$, $E_2 \approx 0.66$ V/nm and $B_\perp = 0$ T) is about 2 K, below which $R_{xy}$ is precisely quantized at $h/e^2$ and $R_{xx}$ vanishes (Fig. 3c, bottom).

The observed state with a nearly quantized $R_{xy}$ and a small but finite $R_{xx}$ is not a Chern insulator: it emerges at $x > 0$ and is compressible (see Fig. S4 for compressibility measurements); it disperses with $B_\perp$ nonlinearly and much more rapidly than a Chern insulator with Chern number one. The state is also not a quantum Hall state: the Landau levels emerge only above $B_{\perp c}$, where the state has already disappeared. Instead, the state is consistent with a Chern metal with chiral edge state transport ($\sigma_{xy} \approx \frac{e^2}{h}$) and a bad metallic bulk ($\sigma_{xx} < \frac{e^2}{h}$). The Chern metal appears right below the Kondo breakdown field $B_{\perp\text{peak}}$ (Fig. 2c) and right next to the Kondo breakdown arc (see Fig. 1d), as revealed by the arc of high $R_{xy}$ in Fig. 2b. It separates two distinct metallic states: the itinerant and the heavy Fermi liquids. It also occurs over an extended range $E_1 < E < E_2$ far beyond the band inversion critical point at $E = E_2$. These results rule out its origin from simply doping the Chern insulator near $E = E_2$. Instead, the results are consistent with the predicted Chern metal phase in the vicinity of a moiré Kondo breakdown(9).

**Chiral edge state transport**
To further support the observation of a Chern metal, we demonstrate chiral edge state transport through the current-induced breakdown and nonlocal transport. Figure 4a shows the differential Hall and longitudinal resistances ($R_{xy}$ and $R_{xx}$) as a function of $x$ and DC bias current $I$ for the same electric field as denoted by the black dashed line in Fig. 2b. The magnetic field is fixed at 2 T. For the Chern metal, the signal peaks at zero bias and is rapidly quenched by increasing the bias current. A line cut at $x = 0.08$ (Fig. 4b, top) shows that the nearly quantized $R_{xy}$ and the $R_{xx}$ dip vanish for current above about 20 nA. In comparison, the resistances show no current dependence except a small current-induced heating effect for a normal metal ($x = 0.25$, bottom). The phenomenology here is similar to that in quantum Hall and quantum anomalous Hall breakdowns(36-38) (see Fig. S5 for the latter). The chiral edge state is quenched by current and the Chern metal becomes a normal metal afterwards.

We also demonstrate the presence of chiral edge states in the Chern metal through nonlocal transport measurements in device 2. Figure 4c shows the schematic of the measurement geometry and an optical micrograph of the device. Specifically, we bias an AC current ($\delta I$) between pins 9 and 8 and measure the voltage ($\delta V$) at different pins along the circumference of the device with reference to pin 8. The total circumference length from pins 1 to 8 is about 20 μm. Figure 4d (left) shows the nonlocal resistance, $R_{NL} \equiv \frac{\delta V}{\delta I}$, as a function of pin index or distance from the current source/drain along the device circumference for a normal metal ($\nu = 1.1$ and $E \approx 0.615$ V/nm), Chern insulator ($\nu = 1$ and $E = E_2 \approx 0.63$ V/nm) and Chern metal ($\nu = 1.07$ and $E \approx 0.615$ V/nm).

For the normal metal ($B_\perp = 0$ T), $R_{NL}$ rapidly decreases with the pin index. This is expected because bulk nonlocal transport is exponentially suppressed with distance from the source-drain location (pins 9 and 8). For the Chern insulator, $R_{NL}$ is independent of the pin index. This is also expected for dissipationless chiral edge transport. But finite nonlocal resistances (about 34 kΩ and 8 kΩ) are observed for both magnetic-field directions ($B_\perp = \pm 2$ T). This arises from the finite contact resistance (8 kΩ) in the three-terminal measurements (at $B_\perp = -2$ T). The equilibrium current for the chiral edge state is clockwise/counterclockwise for two opposite magnetic-field polarities, which results in a difference of $\frac{h}{e^2}$ in $R_{NL}$, as observed in the experiment. For the Chern metal at $B_\perp = 2$ T, $R_{NL}$ decreases slowly with the pin index, exhibiting a dependence intermediate between that of the normal metal and the Chern insulator. (The dependence becomes that of the normal metal at bias currents beyond the chiral edge state breakdown, Fig. S6.) The dependence also becomes that of the normal metal when the polarity of the magnetic field is switched.

The results above support the presence of chiral edge state transport in the Chern metal. The chiral edge state, together with a bad metallic bulk ($\sigma_{xx} < \frac{e^2}{h}$), allows the Chern metal at $B_\perp = 2$ T to resist the strong suppression of nonlocal transport seen in the normal metal. However, the presence of bulk transport in the Chern metal limits the range of nonlocality compared to the Chern insulator. When the magnetic-field polarity is switched (i.e. $B_\perp = -2$ T), pins 1 through 8 would be nearly at the same potential (similar to a Chern insulator), thus resulting in a dependence similar to that of the normal metal. We compare the experiment to the simulated nonlocal response of the device using the Landauer-Büttiker formalism (Fig. 4d, right)(*39*). The transmission matrices for both the chiral edge state and the bad metallic bulk channels (assumed to be equilibrated) are considered and the contact resistances are ignored (Supplementary Materials). The good agreement between the experiment and simulation further supports the presence of a chiral edge state in the Chern metal.

**Magnetization plateau and metamagnetic transition**
Finally, we probe the magnetic response near the Kondo breakdown and the Chern metal phase by magnetic circular dichroism (MCD) spectroscopy(*40*). We spectrally integrate the MCD signal over the moiré exciton resonance of the Mo-layer to obtain a proxy for the spin/valley polarization of the material(*31, 40*) (see Supplementary Materials for details); we also numerically evaluate the differential MCD, $\frac{dMCD}{dB_\perp}$, as a proxy for the differential magnetic susceptibility(*40*). Figures 5a and 5b show the integrated MCD (or simply MCD) and $\frac{dMCD}{dB_\perp}$, respectively, as a function of $B_\perp$ at $T = 1.6$ K and varying doping densities (along the black dashed line in Fig. 2b). Because of the higher temperature compared to transport measurements, the MCD results exhibit stronger thermal broadening effects.

In all cases, the MCD first increases linearly with $B_\perp$ and saturates at high $B_\perp$. For $0.1 \lesssim x \lesssim 0.3$, a MCD plateau, which corresponds to a dip in $\frac{dMCD}{dB_\perp}$, is observed at intermediate fields $2 \lesssim B_\perp \lesssim 5$ T. Moreover, a $\frac{dMCD}{dB_\perp}$ peak near the Kondo breakdown, just before magnetic saturation, is also observed at nearly all doping densities. In Fig. 2c and 5b, we summarize the doping dependence of the magnetic fields at the differential susceptibility dips and peaks ($B_{\perp\text{dip}}$ and $B_{\perp\text{peak}}$, respectively). $B_{\perp\text{dip}}$ and $B_{\perp\text{peak}}$ follow closely the Chern metal phase and the Kondo breakdown. The observed differential susceptibility dip suggests the existence of a spin gap (or a bound state for spins) within the Chern metal phase, while the differential susceptibility peak indicates a metamagnetic transition at the Kondo breakdown.

**Conclusion and outlook**
Through a series of magneto- and nonlocal transport measurements, we demonstrate the emergence of a Chern metal phase near the magnetic field-induced Kondo breakdown in angle-aligned MoTe$_2$/WSe$_2$ moiré bilayers. MCD measurements further suggest the presence of a spin gap for the Chern metal phase and a metamagnetic transition at the Kondo breakdown. Although the results are largely consistent with the predicted Chern metal phase in Ref. (*9*), many open questions remain. First, the observation of a Chern metal arc in the range $E_1 < E < E_2$ (Fig. 2b) suggests the emergence of "manybody" Chern bands preceding the "single-particle" band inversion at $E = E_2$ (Fig. 1c). The close proximity of the Chern metal arc to the Kondo breakdown arc suggests that the Kondo interaction is responsible for the formation of these Chern bands. However, exactly how these Chern bands form near Kondo breakdown remains an open question. Second, the observed bad metallic bulk ($\sigma_{xx} < e^2/h$) is crucial for the experimental manifestation of the Chern metal phase (or else the chiral edge states would be shorted by a good metallic bulk with $\sigma_{xx} \gg e^2/h$). The emergence of a bad metallic bulk suggests strong carrier scattering and/or large quasiparticle effective mass in the vicinity of the Kondo breakdown. This and the metamagnetic transition at $B_{\perp\text{peak}}$ revealed by MCD suggest the field-induced Kondo breakdown is a quantum phase transition rather than a smooth crossover(*34*). Further studies are required to determine whether the transition is abrupt or continuous. Third, the observed magnetization plateau before magnetic saturation is reminiscent of predictions of a spin polaron phase (a bound state of an itinerant carrier and a spin flip) in similar settings as in this study(*41-44*). The possible connection between the Chern metal and the spin polaron phase deserves further investigation. Lastly, we note that the Chern metal phase predicted by the parton mean-field calculations in Ref. (*9*) occurs near a zero-magnetic-field Kondo breakdown in TMD moiré Kondo lattices; the Chern metal is originated from the Kondo coupling between an underlying chiral spin liquid and itinerant carriers. The experimental realization of a zero-field Kondo breakdown in moiré materials is an interesting future direction.

**Acknowledgment**
We thank L. Fu, D. Guerci, A. Kumar, A. Millis, J. Pixley, A. Potter, Q. Si, and S. Todadri for many helpful discussions.

**Author contributions**
W.Z. and Y.Z. fabricated the devices. W.Z. and Y.Z. performed the electrical transport measurements and analyzed the data with the help of B.S. and H.Z. Z.T., W.Z., and P.K. performed the optical measurements. Y.Z. and Z.X. performed the compressibility measurements. K.W. and T.T. grew the bulk hBN crystals. W.Z., J.S., and K.F.M. designed the scientific objectives and oversaw the project. All authors discussed the results and commented on the manuscript.

# Figures

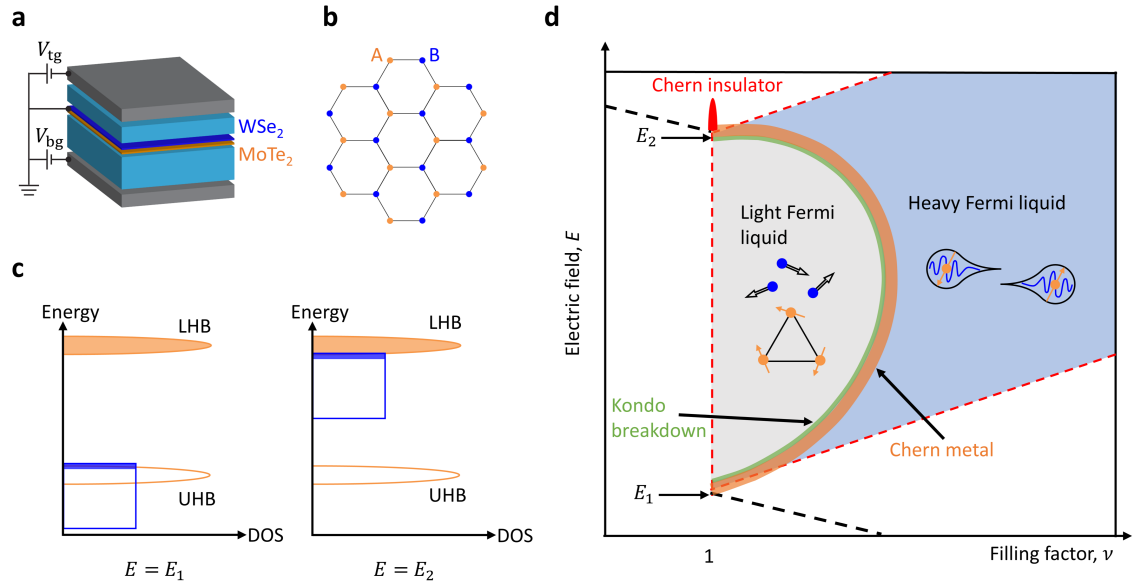

**Figure 1. Gate-tunable moiré Kondo lattice**. **a**, Schematic of a dual-gated $MoTe_2/WSe_2$ moiré bilayer device. The gates consist of hBN dielectrics (light blue) and few-layer graphite electrodes (grey) with $V_{tg}$ and $V_{bg}$ denoting the top and bottom gate voltages, respectively. **b**, The hole Wannier orbitals from the two layers form the sublattices of a honeycomb moiré lattice. **c**, Schematic energy dependence of the electronic density of states (DOS) at total hole filling factor $\nu = 1$ and critical out-of-plane electric field $E_1$ (left) and $E_2$ (right). A Mott gap separates the filled lower Hubbard band (LHB) and the empty upper Hubbard band (UHB) of the Mo-layer. The maximum of the dispersive hole band of the W-layer is aligned with the UHB maximum at $E = E_1$ and with the LHB minimum at $E = E_2$. **d**, Schematic phase diagram in $\nu$ and $E$. The red dashed lines enclose the Kondo lattice region ($\nu > 1$ and $E_1 < E < E_2$); the critical fields increase linearly with $\nu$. The green arc marks the Kondo breakdown boundary separating a heavy and a light Fermi liquid (with arrows denoting the spins) under a perpendicular magnetic field. The orange arc denotes the Chern metal phase right next to the Kondo breakdown arc. The red region at $\nu = 1$ and $E \approx E_2$ marks the Chern insulator. In **a-d**, holes or hole bands from the Mo- and W-layer are shown in orange and blue, respectively.

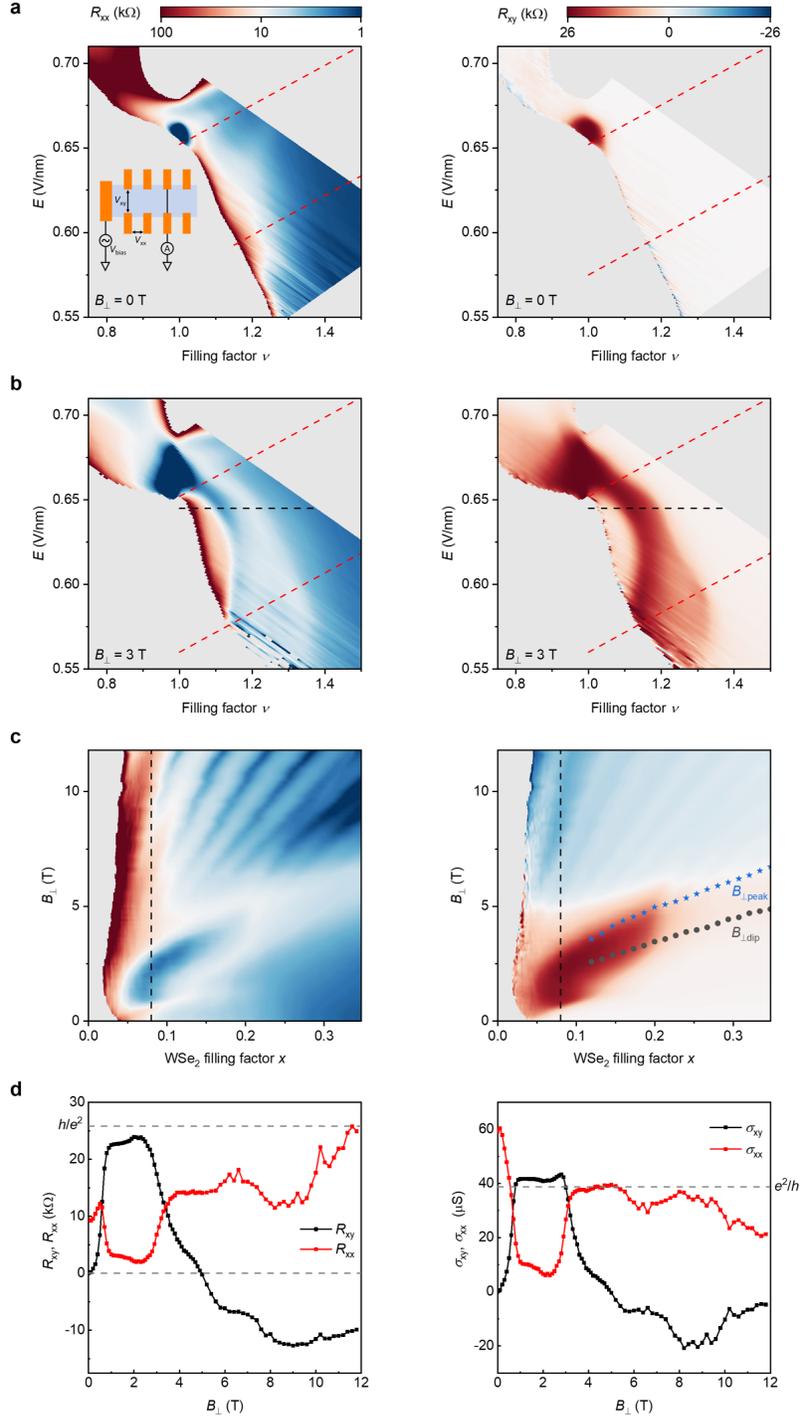

**Figure 2. Emergence of Chern metal**. **a,b**, Longitudinal ($R_{xx}$, left) and Hall ($R_{xy}$, right) resistances as a function of $\nu$ and $E$ under out-of-plane magnetic field $B_\perp = 0$ T (**a**) and 3 T (**b**). The inset in **a** shows the measurement geometry. Between the red dashed lines is the Kondo lattice region. **c**, $R_{xx}$ (left) and $R_{xy}$ (right) as a function of the hole-filling factor

of the W-layer $x$ and $B_\perp$ for a representative electric field from the Kondo lattice region (black dashed line in **b**). Symbols: the Kondo breakdown critical field $B_{\perp\text{peak}}$ and the field at the differential susceptibility minimum $B_{\perp\text{dip}}$ from Fig. **5b**. **d**, Magnetic-field dependence of $R_{xx}$ and $R_{xy}$ (left) and of the corresponding longitudinal ($\sigma_{xx}$) and Hall ($\sigma_{xy}$) conductivities (right) at $x = 0.08$. The Chern metal manifests a nearly quantized $R_{xy} \approx \frac{h}{e^2}$ (or $\sigma_{xy} \approx \frac{e^2}{h}$) and a strongly suppressed $R_{xx} \ll \frac{h}{e^2}$ (or $\sigma_{xx} \ll \frac{e^2}{h}$). The grey-shaded regions in **a-c** are not accessible; they are either beyond the gate limits or too resistive for reliable four-terminal measurements. The sample (lattice) temperature is 10 mK.

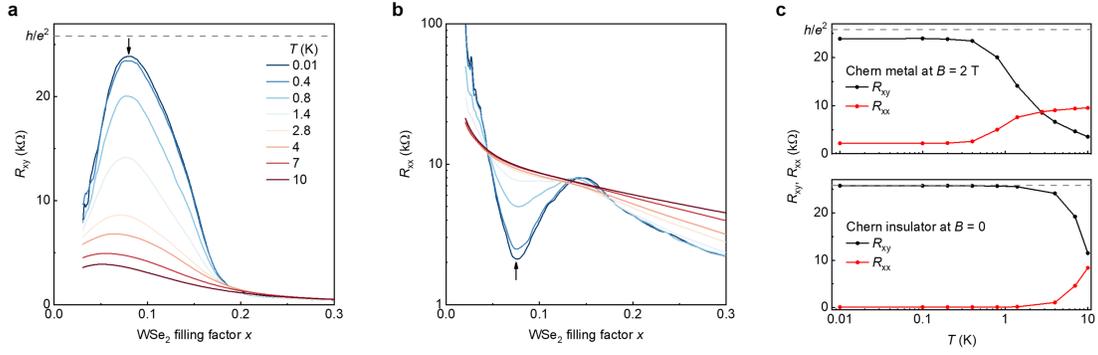

**Figure 3. Temperature dependence of the Chern metal**. **a,b**, Filling factor dependence of $R_{xy}$ (**a**) and $R_{xx}$ (**b**) at varying temperatures (along the black dashed line in Fig. **2b** and at $B_\perp = 2$ T). The arrows denote the Chern metal near $x = 0.08$, exhibiting a dip in $R_{xx}$ and a peak in $R_{xy}$. **c**, Temperature dependence of $R_{xx}$ and $R_{xy}$ for the Chern metal at $x = 0.08$ and $B_\perp = 2$ T (top) and for the Chern insulator at $\nu = 1$ and $B_\perp = 0$ T (bottom). The Chern metal (insulator) displays a nearly (precisely) quantized $R_{xy}$ and a small but finite (vanishing) $R_{xx}$.

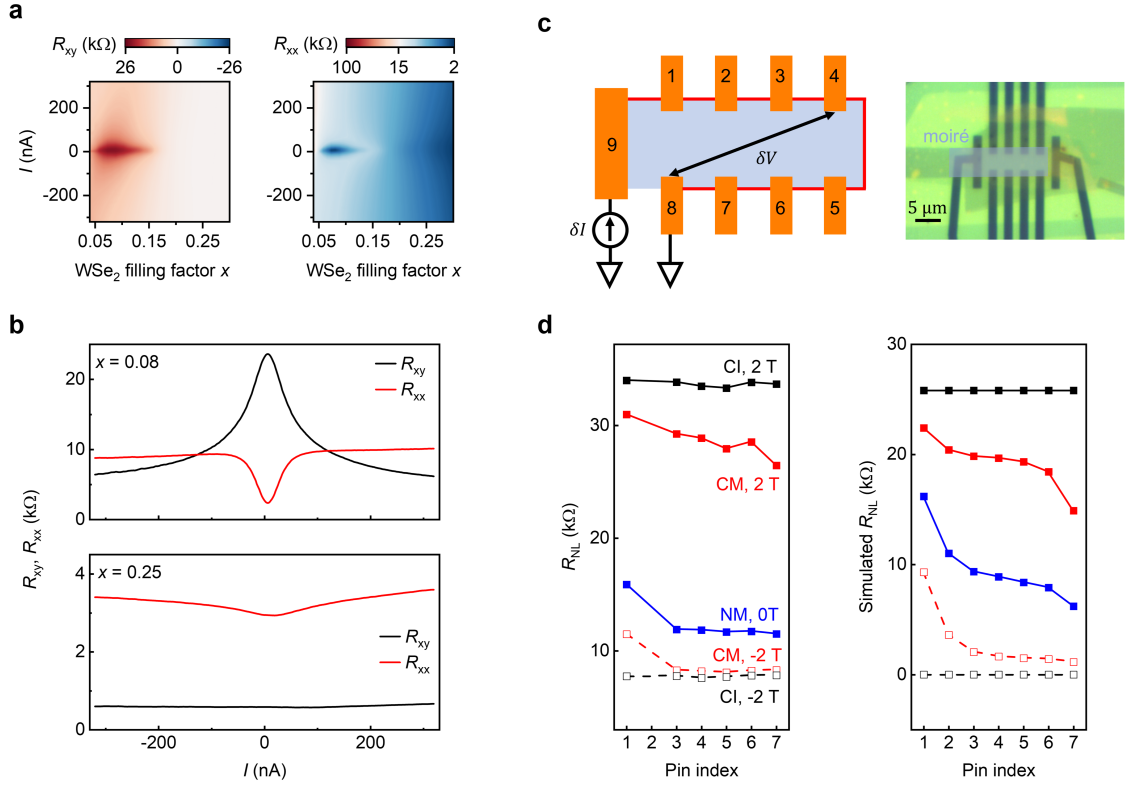

**Figure 4. Chiral edge state transport**. **a**, Differential Hall resistance $R_{xy}$ (left) and longitudinal resistance $R_{xx}$ (right) as a function of the filling factor in the W-layer $x$ and DC bias current $I$ (along the black dashed line in Fig. **2b**, $B_\perp = 2$ T). **b**, Line cuts of **a** at $x = 0.08$ (top, Chern metal) and $x = 0.25$ (bottom). Only the Chern metal exhibits quantum anomalous Hall breakdown. **c**, Schematic for the nonlocal transport measurement geometry (left) corresponding to the optical micrograph of device 2 (right). The uniform moiré region is shaded in blue. An AC current $\delta I$ is biased between pin 9 and 8, and the nonlocal resistance $R_{NL} = \frac{\delta V}{\delta I}$ is determined by probing voltage $\delta V$ at different pins along the device circumference (red line) with reference to pin 8. **d**, Nonlocal resistance of device 2 (experiment, left; simulation, right) as a function of pin index for the Chern insulator (CI) and the Chern metal (CM) at $B_\perp = \pm 2$ T and for a normal metal (NM) at $B_\perp = 0$ T. The nonlocal resistance is independent of the pin index for the Chern insulator ($\nu = 1$ and $E \approx 0.63$ V/nm) and decays rapidly for the normal metal ($\nu = 1.07$ and $E \approx 0.615$ V/nm). It decays slowly (quickly) at $B_\perp = +2$ T ($-2$ T) for the Chern metal ($\nu = 1.07$ and $E \approx 0.615$ V/nm). The sample (lattice) temperature is 10 mK.

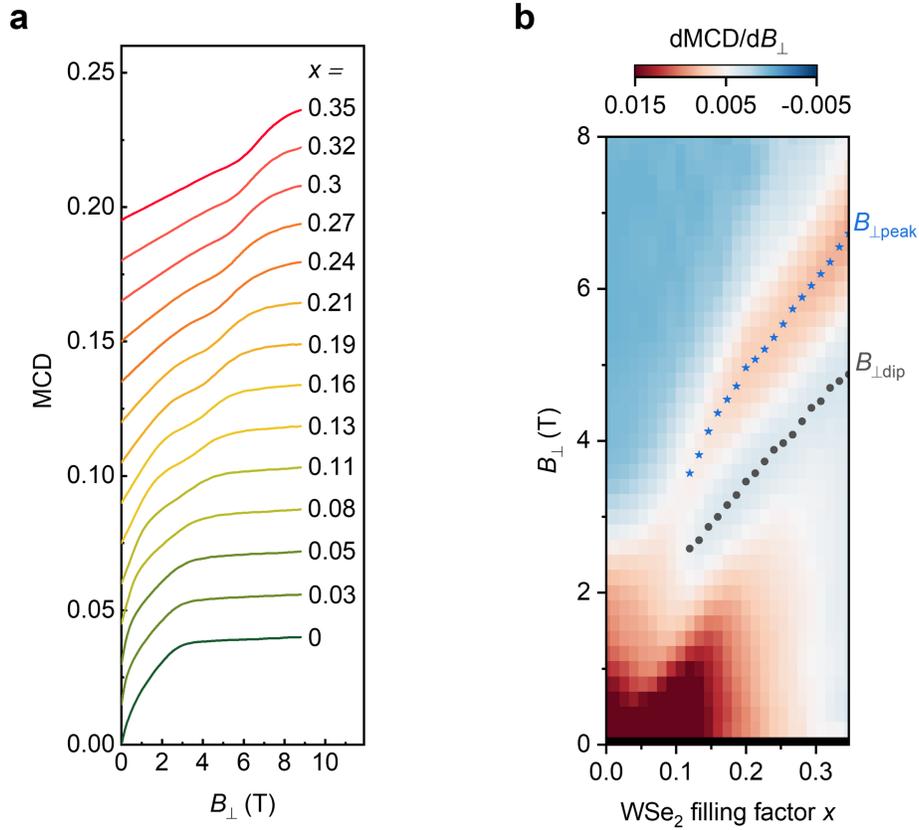

**Figure 5. MCD plateau and metamagnetic transition near the Kondo breakdown. a**, Magnetic-field dependence of spectrally integrated magnetic circular dichroism (MCD) at varying filling factors of the W-layer. It exhibits a plateau followed by a rapid increase with increasing magnetic field for $0.1 \lesssim x \lesssim 0.3$. The curves at different fillings are vertically displaced for clarity. **b**, Differential MCD ($\frac{dMCD}{dB_\perp}$) as a function of $x$ and $B_\perp$. The dots and stars denote, respectively, the dips and peaks of the differential MCD (also shown in Fig. **2c**). The Chern metal phase coincides with the dip, and the Kondo breakdown coincides with the peak. The electric field is fixed along the black dashed line in Fig. **2b** and the sample temperature is 1.6 K.

**Supplementary Materials for**
**"Emergence of Chern metal in a moiré Kondo lattice"**

Wenjin Zhao[*†], Zui Tao[*], Yichi Zhang[*], Bowen Shen, Zhongdong Han, Patrick Knüppel, Yihang Zeng, Zhengchao Xia, Kenji Watanabe, Takashi Taniguchi, Jie Shan[†], Kin Fai Mak[†]

[*]These authors contributed equally.
[†]Email: wz435@cornell.edu; jie.shan@cornell.edu; kinfai.mak@cornell.edu

**The PDF file includes:**

Materials and Methods
References (*1-4*)
Figs. S1 to S7

## Materials and Methods

### Device fabrication and electrical measurements

We fabricated dual-gate devices of angle-aligned MoTe$_2$/WSe$_2$ bilayers using the layer-by-layer dry transfer technique, as detailed in Refs. (*10, 23*). In short, both the bottom and top gates of the devices were made of hexagonal boron nitride (hBN) and few-layer graphite. The top gate hBN has a thickness of about 4 nm to facilitate the application of a large electric field. We used 5-nm-thick Pt as the contact electrodes to the MoTe$_2$/WSe$_2$ bilayers. The two gates allow independent control of the hole doping density and the out-of-plane electric field. The electrical contacts are turned on under high electric fields.

Electrical transport measurements were conducted in a dilution refrigerator (Bluefors LD250) equipped with a 12 T superconducting magnet. Standard lock-in techniques were employed to measure the differential resistances; a voltage pre-amplifier with 100-MΩ impedance was used to measure the sample resistance up to about 10 MΩ. Our devices exhibit finite longitudinal-transverse coupling. We symmetrized and anti-symmetrized the measured $R_{xx}$ and $R_{xy}$ under positive and negative magnetic fields, respectively, to obtain the longitudinal and Hall resistances.

### Optical measurements

We performed reflective magnetic circular dichroism (MCD) spectroscopy down to 1.6 K in a closed-cycle $^4$He cryostat (Attocube, Attodry 2100). Details of the measurement have been reported in Ref. (*40*). The light source was a superluminescent diode centered around 1070 nm, and the detector was a liquid nitrogen-cooled InGaAs array sensor. A microscope objective with a numerical aperture of 0.8 was used to focus light onto the devices with a spot of about 1.5 μm in diameter. To prevent heating or photo-doping the sample, the incident light intensity was kept below 20 nW/μm². The MCD spectrum is defined as the difference in reflectance between the left- and right-handed circularly polarized light ($I^+$ and $I^-$, respectively), normalized by their sum: $\frac{I^+ - I^-}{I^+ + I^-}$. We observed a strong enhancement of MCD near the attractive polaron resonance of MoTe$_2$ (Fig. S7). The absolute value of the MCD was integrated over a spectral window ranging from 1.117 to 1.138 eV that covers the attractive polaron resonance.

### Simulation

We used the Landauer–Büttiker formula(*39*) to simulate the nonlocal resistance of the Hall bar device shown in Fig. 4c:

$$I_i = \frac{e^2}{h} \sum_j (T_{ji} V_i - T_{ij} V_j).$$

Here $I_i$ is the net current leaving terminal $i$, $T_{ij}$ is the transmission probability from terminal $j$ to terminal $i$, and $V_i$ is the voltage at terminal $i$. We treated the transmission matrix as a sum of the contributions from the chiral edge state $T_{edge}$ and from the bulk $T_{bulk}$. The chemical potentials of the edge and bulk states are equilibrated at each terminal.

The edge state transmission matrix for the measurement geometry in Fig. 4c is a $9 \times 9$ matrix, taking the form below when $B_\perp > 0$:

$$T_{edge} = \begin{pmatrix} 0 & 0 & 0 & 0 & 0 & 0 & 0 & 0 & 1 \\ 1 & 0 & 0 & 0 & 0 & 0 & 0 & 0 & 0 \\ 0 & 1 & 0 & 0 & 0 & 0 & 0 & 0 & 0 \\ 0 & 0 & 1 & 0 & 0 & 0 & 0 & 0 & 0 \\ 0 & 0 & 0 & 1 & 0 & 0 & 0 & 0 & 0 \\ 0 & 0 & 0 & 0 & 1 & 0 & 0 & 0 & 0 \\ 0 & 0 & 0 & 0 & 0 & 1 & 0 & 0 & 0 \\ 0 & 0 & 0 & 0 & 0 & 0 & 1 & 0 & 0 \\ 0 & 0 & 0 & 0 & 0 & 0 & 0 & 1 & 0 \end{pmatrix}.$$

Assuming a uniform bulk conductivity $\sigma_{xx}$, we used COMSOL to simulate the bulk transmission matrix

$$T_{bulk} = \sigma_{xx} \frac{h}{e^2} \begin{pmatrix} 0 & 0.859 & 0.0112 & 6 \times 10^{-4} & 6 \times 10^{-4} & 0.01 & 0.122 & 0.303 & 1.114 \\ 0.859 & 0 & 0.862 & 0.0119 & 0.108 & 0.125 & 0.344 & 0.123 & 0.284 \\ 0.0112 & 0.862 & 0 & 0.873 & 0.136 & 0.344 & 0.125 & 0.01 & 1.6 \times 10^{-3} \\ 6 \times 10^{-4} & 0.0119 & 0.873 & 0 & 0.564 & 0.136 & 0.01 & 6 \times 10^{-4} & 9 \times 10^{-5} \\ 6 \times 10^{-4} & 0.108 & 0.136 & 0.564 & 0 & 0.873 & 0.012 & 6 \times 10^{-4} & 9 \times 10^{-5} \\ 0.01 & 0.125 & 0.344 & 0.136 & 0.873 & 0 & 0.862 & 0.0112 & 0.0016 \\ 0.122 & 0.344 & 0.125 & 0.01 & 0.012 & 0.862 & 0 & 0.859 & 0.0284 \\ 0.303 & 0.123 & 0.01 & 6 \times 10^{-4} & 6 \times 10^{-4} & 0.0112 & 0.859 & 0 & 1.114 \\ 1.114 & 0.284 & 1.6 \times 10^{-3} & 9 \times 10^{-5} & 9 \times 10^{-5} & 0.0016 & 0.0284 & 1.114 & 0 \end{pmatrix}.$$

We calculated the nonlocal resistance for the following cases:
1) A Chern insulator with $\sigma_{xx} = 0$;
2) A Chern metal with a bad metallic bulk conductivity $\sigma_{xx} < \frac{e^2}{h}$;
3) A normal metal with $\sigma_{xx} \gg \frac{e^2}{h}$.

The calculated nonlocal resistances show good agreement with the experimental results in Fig. 4d. Note that contact resistance is ignored in the calculations. We chose $\sigma_{xx} \approx 0.6 \frac{e^2}{h}$ for the Chern metal case to obtain a good match with experiment. The value is also in reasonable agreement with the measured $\sigma_{xx}$ in Fig. 2d. When the chirality of the edge state flips under $B_\perp < 0$, the edge state transmission matrix becomes:

$$T'_{edge} = \begin{pmatrix} 0 & 1 & 0 & 0 & 0 & 0 & 0 & 0 & 0 \\ 0 & 0 & 1 & 0 & 0 & 0 & 0 & 0 & 0 \\ 0 & 0 & 0 & 1 & 0 & 0 & 0 & 0 & 0 \\ 0 & 0 & 0 & 0 & 1 & 0 & 0 & 0 & 0 \\ 0 & 0 & 0 & 0 & 0 & 1 & 0 & 0 & 0 \\ 0 & 0 & 0 & 0 & 0 & 0 & 1 & 0 & 0 \\ 0 & 0 & 0 & 0 & 0 & 0 & 0 & 1 & 0 \\ 0 & 0 & 0 & 0 & 0 & 0 & 0 & 0 & 1 \\ 1 & 0 & 0 & 0 & 0 & 0 & 0 & 0 & 0 \end{pmatrix}.$$

The calculated nonlocal resistances are also in good agreement with experiment for $B_\perp < 0$ (Fig. 4d).

**Supplementary Figures**

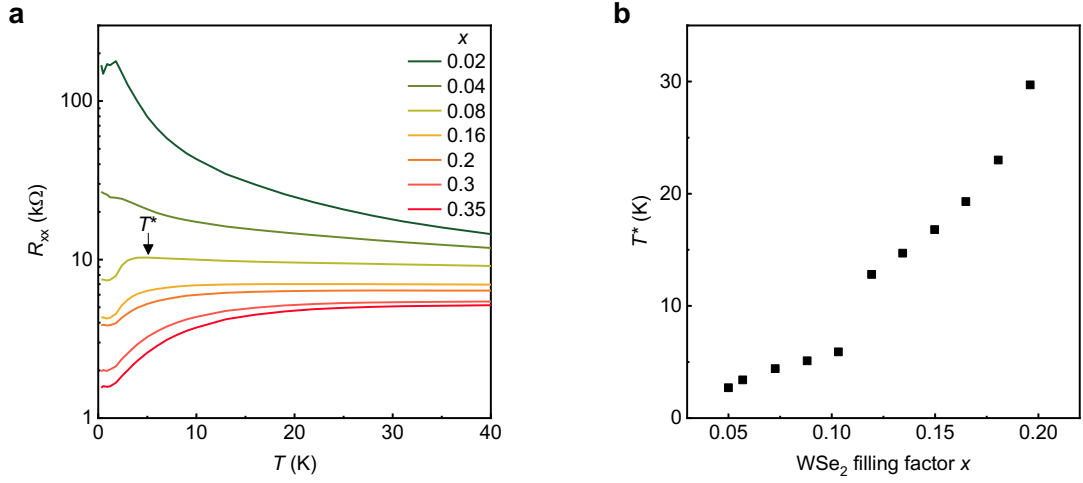

**Figure S1. Gate-tunable Kondo temperature. a**, Temperature-dependent $R_{xx}$ at varying doping densities for $\nu = 1 + x$ (along the black dashed line in Fig. 2b at $B_\perp = 0$ T). The arrow marks the Kondo temperature $T^*$, below which coherent transport emerges. **b**, Dependence of $T^*$ on WSe$_2$ filling factor for $\nu = 1 + x$. $T^*$ vanishes near $x = 0.05$.

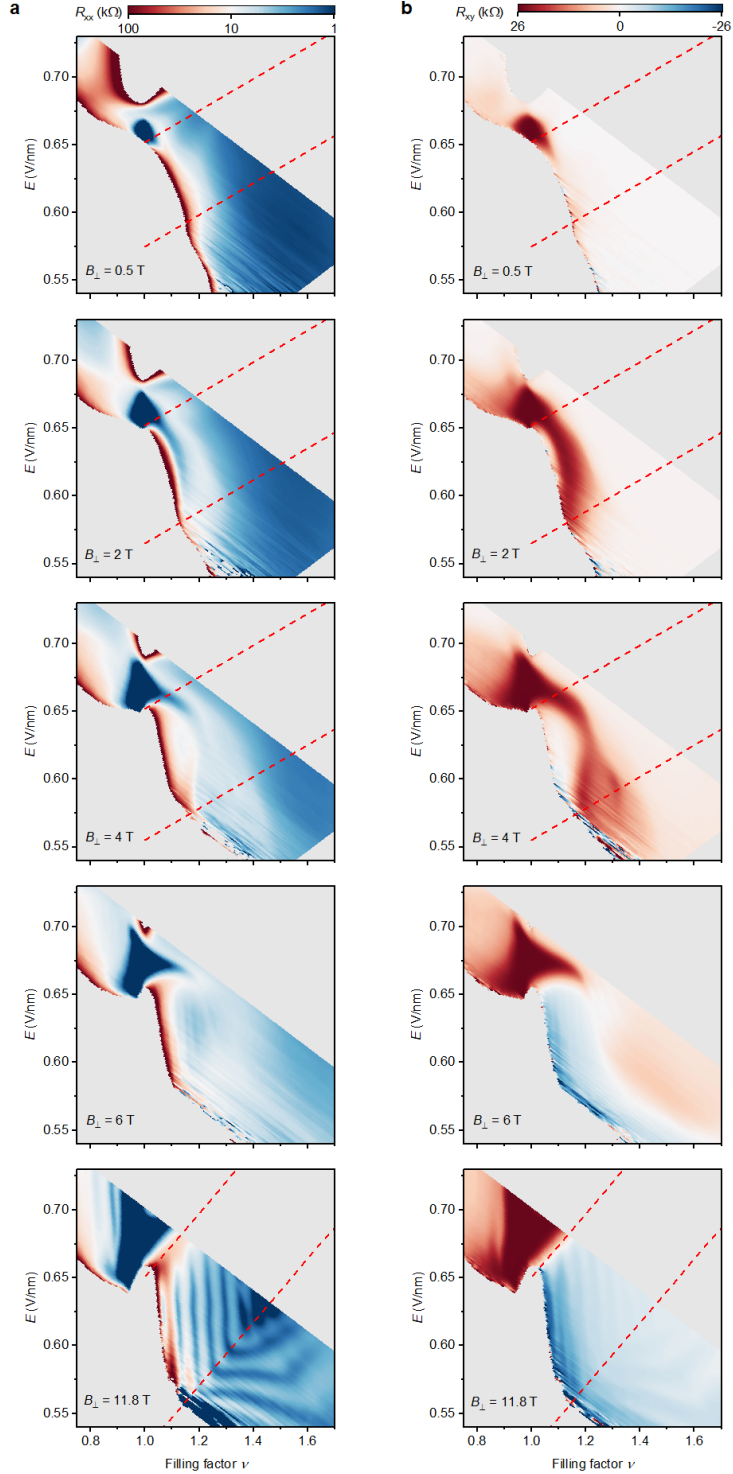

**Figure S2.** $R_{xx}$ (a) and $R_{xy}$ (b) as a function of $\nu$ and $E$ at different magnetic fields. The Chern metal arc near the Kondo breakdown emerges at moderate magnetic fields and disappears at high fields. The red-dashed lines denote the Kondo lattice region. The grey-shaded areas represent regions either beyond the gate limits or with resistances too large for reliable four-terminal measurements. The sample temperature is at $T = 10$ mK (lattice temperature).

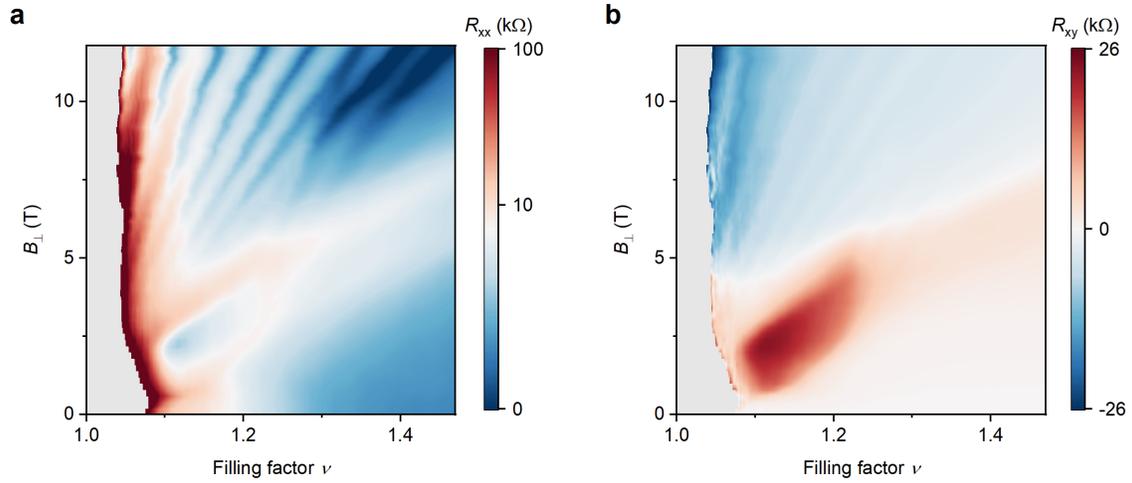

**Figure S3.** $R_{xx}$ (a) and $R_{xy}$ (b) as a function of $\nu$ and $B_\perp$ at $E = 0.625$ V/nm. A Landau fan emerges beyond the Kondo breakdown field $B_{\perp\text{peak}}$. A Chern metal phase with a $R_{xx}$ dip and $R_{xy}$ peak is observed at $1.07 \lesssim \nu \lesssim 1.23$ near but before the Kondo breakdown. The sample temperature is at $T = 10$ mK (lattice temperature).

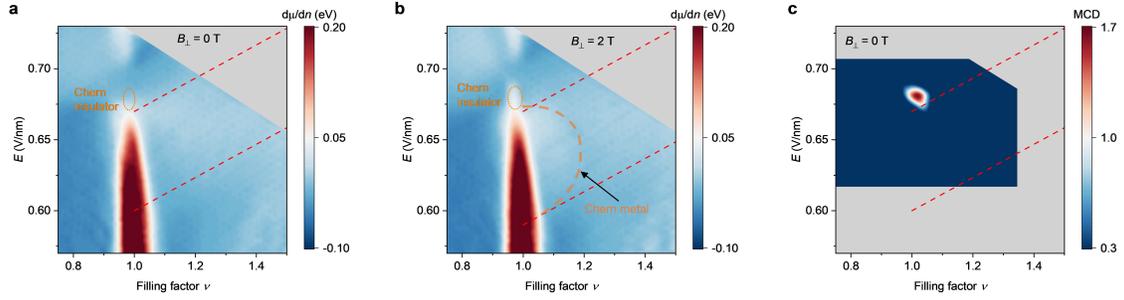

**Figure S4. Electronic incompressibility of MoTe$_2$/WSe$_2$ moiré bilayer. a, b**, The electronic incompressibility as a function of $\nu$ and $E$ at $T = 1.6$ K and $B_\perp = 0$ T (**a**) and 2 T (**b**). The red-dashed lines denote the Kondo lattice region; the Chern insulator is circled; the orange-dashed line in **b** locates the Chern metal arc. An incompressible state is observed only at $\nu = 1$. The states near the Kondo breakdown arc at $B_\perp = 2$ T are compressible. **c,** Spontaneous MCD (at $B_\perp = 0$ T) as a function of $\nu$ and $E$ at $T = 1.6$ K. Only the Chern insulating state at $\nu = 1$ shows a spontaneous MCD hot spot.

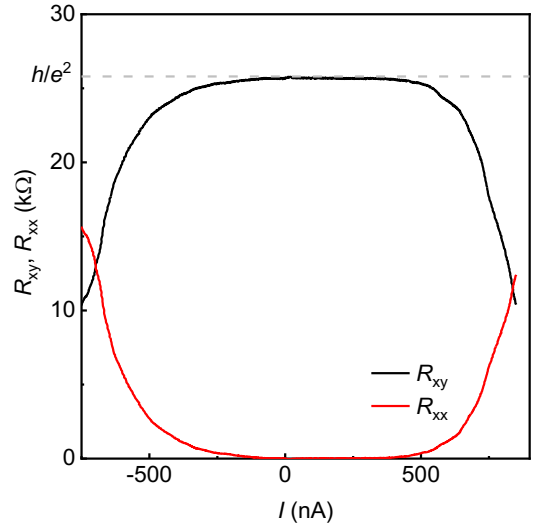

**Figure S5. Quantum anomalous Hall breakdown.** Differential longitudinal ($R_{xx}$) and Hall ($R_{xy}$) resistances as a function of the DC bias current for the Chern insulator at $\nu = 1$ and $B_\perp = 2\,\text{T}$ ($T = 10\,\text{mK}$). The quantum anomalous Hall breakdown indicated by the deviation of $R_{xy}$ from the quantized value (dashed line) and of $R_{xx}$ from 0 happens at a much higher critical current compared to the Chern metal phase (Fig. 4b).

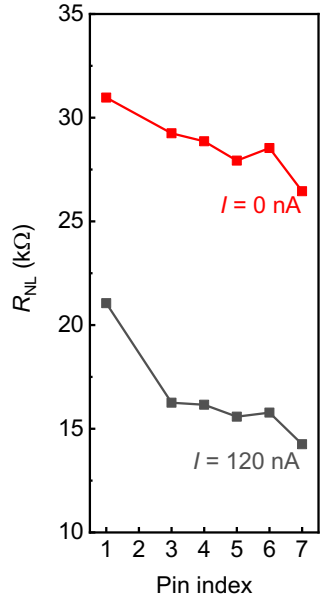

**Figure S6. Additional nonlocal transport measurements.** Nonlocal resistance $R_{NL}$ for the Chern metal phase at $B_\perp = 2$ T and $T = 10$ mK (lattice temperature) as a function of pin index at different bias currents. A slow decay for the Chern metal at $I = 0$ nA turns to a fast decay for a normal metal at $I = 120$ nA due to quantum anomalous Hall breakdown of the chiral edge states at high bias currents.

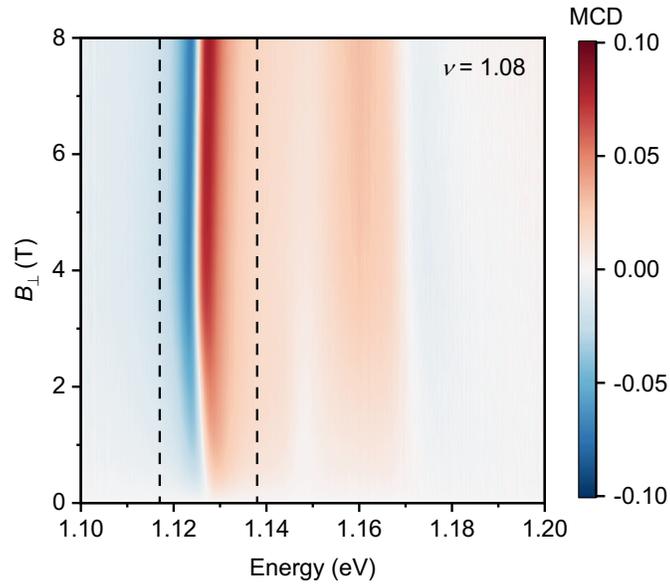

**Figure S7. Magnetic field dependent MCD spectrum near the MoTe$_2$ moiré exciton resonance.** The filling factor is $v = 1.08$; the electric field is $E = 0.645$ V/nm; the sample temperature is $T = 1.6$ K. The dashed lines bound the spectral range for the MCD integration; the range covers the fundamental moiré exciton resonance of the Mo-layer, where resonantly enhanced MCD response is observed.